\UseRawInputEncoding
\documentclass[11pt]{spie}  

\usepackage[utf8]{inputenc}
\usepackage{amsmath,amsfonts,amssymb}
\usepackage{comment}
\usepackage{graphicx}
\usepackage[colorlinks=true, allcolors=blue]{hyperref}

\title{Segmenting Cardiac Muscle Z-disks with Deep Neural Networks}

\author[1,2]{Mihaela Croitor Ibrahim}
\author[1]{Nishant Ravikumar}
\author[3,4]{Alistair Curd}
\author[1]{Joanna Leng}
\author[1]{Oliver Umney}
\author[3]{Michelle Peckham}
\affil[1]{School of Computing, University of Leeds, UK}
\affil[2]{School of Computing, University College London, UK}
\affil[3]{School of Molecular and Cellular Biology, University of Leeds, UK
}
\affil[4]{Division of Pathology and Data Analytics, University of Leeds, UK}

\begin{document} 

\maketitle
\begin{abstract}
Z-disks are complex structures that delineate repeating sarcomeres in striated muscle. They play significant roles in cardiomyocytes such as providing mechanical stability for the contracting sarcomere, cell signalling and autophagy. Changes in Z-disk architecture have been associated with impaired cardiac function. Hence, there is a strong need to create tools to segment Z-disks from microscopy images, that overcome traditional limitations such as variability in image brightness and staining technique.
In this study, we apply deep learning based segmentation models to extract Z-disks in images of striated muscle tissue. We leverage a novel Airyscan confocal dataset, which comprises high resolution images of Z-disks of healthy heart tissue, stained with Affimers for specific Z-disk proteins. We employed an interactive labelling tool, Ilastik to obtain ground truth segmentation masks and use the resulting data set to train and evaluate the performance of several state-of-the-art segmentation networks.
On the test set, UNet++ achieves best segmentation performance for Z-disks in cardiomyocytes, with an average Dice score of 0.91 and outperforms other established segmentation methods including UNet, FPN, DeepLabv3+ and pix2pix. However, pix2pix demonstrates improved generalisation, when tested on an additional dataset of cardiomyocytes with a titin mutation. This is the first study to demonstrate that automated machine learning-based segmentation approaches may be used effectively to segment Z-disks in confocal microscopy images. Automated segmentation approaches and predicted segmentation masks could be used to derive morphological features of Z-disks (e.g. width and orientation), and subsequently, to quantify disease-related changes to cardiac microstructure.
\end{abstract}


\section{INTRODUCTION}
\label{sec:intro}  

\textbf{Importance of Z-disks.} Cardiac muscle, like skeletal muscle, has a striated appearance caused by recurrent arrangements of thick and thin filaments in sarcomeres \cite{sweeney_muscle_2018,pinnell_cardiac_2007}. These are the basic contractile units of the muscle, and are delineated by thin regions called Z-disks \cite{sweeney_muscle_2018,hopkins_skeletal_2006}. Z-disks are complex structures, made of actin filaments cross-linked by alpha-actinin molecules and several other proteins including titin, and nebulin \cite{luther_vertebrate_2009,khadangi_automated_2019,morris_striated_2020}. Z-disks play a crucial role in regulating healthy contractile movement, by anchoring sarcomeric proteins \cite{luther_vertebrate_2009} . Recently, several other roles have been attributed to Z-disks including intracellular signalling, mechanotransduction, mechanosensation and protein synthesis and degradation \cite{knoll_sarcomeric_2011}. Any changes in structure and topology of the Z-disks could disrupt healthy cardiac function by impairing contraction and reducing its blood pumping capability \cite{sequeira_physiological_2014,knoll_sarcomeric_2011}. Mutations in  Z-disk proteins lead to  a range of cardiac diseases including dilated cardiomyopathy (DCM), hypertrophic cardiomyopathy (HCM), muscular dystrophy and myofibrillar myopathy \cite{wadmore_role_2021,mcnally_genetic_2013}. \\ 
\textbf{Z-disk imaging.} Historically, researchers have evaluated sarcomere structure by staining for $\alpha$-actinin proteins, which are mainly located in the Z-disks \cite{ehler_myofibrillogenesis_1999,drew_metrics_2015,drew_multiscale_2016,noureddine_structural_2023}. Antibodies are the most common choice of binder when locating this protein  \cite{saide_characterization_1989,young_molecular_1998,noureddine_structural_2023}, though they have limited penetration because of their large size. Recently, novel binders termed Affimers were isolated for specific  Z-disk proteins including alpha-actinin, ZASP and titin \cite{parker_affimers_2023}. In combination with super-resolution imaging techniques such as Airyscan confocal and STED microscopy, Affimers showed improved resolution and revealed new structural details in the Z-disks \cite{parker_affimers_2023}, enabling investigation of cardiomyocyte structure and identification of disease phenotypes.\\ 
\textbf{Segmentation of Z-disks.} Quantitative assessment of Z-disk morphology is essential to explain disease mechanisms in cells and requires delineation/segmentation of Z-disks from images of cardiomyocytes. Manually tracing the contours of Z-disks requires expertise, is time-consuming, and is subject to intra-/inter-observer variability. A step towards automation was taken through the development of methods like ZlineDetection \cite{morris_striated_2020}, a Matlab implementation which extracts Z-disks via an image processing-based algorithm. However, this lacks generalisability as it depends on thresholding pixel values, which can vary greatly with different staining techniques and image quality. An automated workflow, involving image pre-processing, segmentation and refinement, using contrast sketching, Gaussian kernels and Sobel operators was previously proposed \cite{khadangi_automated_2019}.
Although effective, this method was outperformed by an interactive UNet-based segmentation tool integrated in the bioimage analysis software Ilastik, which was trained on only 8 sparse annotated images \cite{khadangi_automated_2019}.
Despite achieving ground-breaking results in various medical imaging applications, deep learning-based segmentation is still relatively unexplored for segmentation of Z-disks. This is attributed to the lack of suitable training data sets. 
In this study, we utilise the Ilastik software, to interactively label a dataset of cardiomyocytes stained with Affimers and imaged with Airyscan confocal microscopy We explore a wide range of popular segmentation techniques including UNet, UNet++, DeepLabV3+ and FPN (feature pyramid network) and one based on pix2pix \cite{isola_image--image_2018}, a conditional generative adversarial network (cGAN). \\
\textbf{Contributions.} (i) This is the first study to investigate automated deep learning-based segmentation of  Z-disks in striated muscle tissue; (ii) We present a systematic evaluation of several state-of-the-art segmentation techniques: UNet++, DeepLabV3, FPN on a novel dataset.

\section{Methodology}

\textbf{Dataset.} The images used in this work were generated from staining cryo-sections of human cardiac tissue from normal donors as described in Parker, et al.\cite{parker_affimers_2023}. The tissue samples were obtained from the Sydney heart bank with consent obtained from St Vincent's Hospital, Darlinghurst (HREC \#H91/048/1a) and the University of Sydney (HREC \#2016/923), prepared according to a standardised protocol\cite{parker_affimers_2023} and stained using a dye-labelled Affimer to highlight the Z-disk proteins $\alpha$-actinin-2\cite{curd_nanoscale_2021}, Z1Z2 or to ZASP \cite{parker_affimers_2023}, followed by mouse monoclonal antibody to human desmoglein-2 (CCSTEM28; eBiosciences from Thermo
Scientific, 1:200). Sections were imaged using a Zeiss LSM 880 Airyscan confocal, using a x40 objective (NA 1.4) in Airyscan mode. \\
\textbf{Semi-interactive labelling.} The requirement for large annotated datasets remains a bottleneck in utilising supervised Deep Learning effectively for semantic segmentation. This is particularly difficult for histopathology/microscopy images which are high-dimensional, and where expert knowledge is imperative to manually contour structures of interest. This is a limiting factor in the volume of images that can be processed in a timely-manner. To address this challenge, several interactive annotation tools have been created for microscopy data including AnnotatorJ\cite{hollandi_annotatorj_2020}, microSAM and Ilastik\cite{berg_ilastik_2019}. In this study, we used Ilastik Pixel Classification \cite{berg_ilastik_2019}, a pre-trained U-Net model that can be interactively trained using sparse labels to annotate images and curate ground-truth segmentation masks. This bio-image analysis software is user-friendly, stable, and can be customised to certain structures. To reduce processing time, users can interactively select which features are relevant to the task. 

\textbf{Data pre-processing.} The initial dataset contained 130 images of a normal patient heart. 10\% of this data was kept aside for testing. The remaining data was used for training and validation across 3-fold cross-validation experiments. To test generalisability, an additional test set was created containing  10 images of heart tissue with a titin mutation,\cite{parker_affimers_2023}. Each original image of size 2056*2056 pixels was resized to 1024*1024 and cropped into 16 patches 256*256 pixels in size. 

\textbf{Methods.} The study evaluates the performance of established deep segmentation networks for segmenting Z-disks in Airyscan confocal microscopy images. First, we investigated the UNet \cite{ronneberger_u-net_2015}, for segmentation of biomedical images. Its key aspects are an encoder-decoder network architecture, with the introduction of skip connections between corresponding convolution blocks in the encoder and decoder network branches. The encoder downsamples the image to extract high-level information, while the decoder maps the bottleneck features to high-resolution input feature maps. The skip connections between the encoder and the decoder improve information/gradient flow across the network, enabling learning of both local and global contextual information. For the UNet architecture, we utilised a pre-trained VGG11 encoder imported from an established python library, torchvision \cite{torchvision2016}. UNet has inspired several subsequent segmentation networks including UNet++ \cite{azad_medical_2022,yin_u-net-based_2022}. UNet++ aims to reduce the semantic gap between the feature maps of the encoder and decoder using nested, dense skip connections \cite{zhou_unet_2018}. Through deep supervision and efficient aggregation of features from different decoder layers, UNet++ has been demonstrated to improve segmentation performance (relative to a UNet) \cite{yin_u-net-based_2022}. We paired this architecture with a pre-trained Efficient-b4 encoder in this study \cite{torchvision2016}. Another family of models is built around DeepLab \cite{chen_semantic_2016}.
DeepLabV3+, the latest improvement to DeepLab, combines the atrous convolution operations and atrous spatial pyramid pooling operations proposed in DeepLab, with an encoder-decoder architecture \cite{chen_semantic_2016}. DeepLabV3+ introduces depth-wise convolutions to both encoder and decoder, and has been shown to achieve state-of-the-art segmentation performance \cite{chen_encoder-decoder_2018,khan_evaluation_2020}. We paired the DeepLabV3+ architecture with a pre-trained ResNet34 encoder \cite{torchvision2016}. Another network that aims to capture multi-scale contextual information is FPN (feature pyramid network)\cite{lin_feature_2017}. FPN employs a top-down pathway and a bottom-up pathway linked through lateral connections to create a pyramid of features, suitable for object detection and semantic segmentation. Similarly to the DeepLabV3+, we paired this network with a ResNet34 encoder \cite{torchvision2016}.\\ 
Pix2pix is a type of conditional Generative Adversarial Network (cGAN)\cite{isola_image--image_2018} which employs a UNet generator and a patch-based discriminator. The generator and discriminator are trained adversarially to learn the conditional mapping from source to target domain images (i.e. image translation) \cite{goodfellow_generative_2014}. In addition to the adversarial loss \cite{mirza_conditional_2014}, the model often includes a pixel-wise L1 loss \cite{janocha_loss_2017} forcing the generator to not only produce outputs that are realistic but also close to the ground truth. We utilise pix2pix for synthesising segmentation maps given input images in this study and explore two implementations: (i) with L1 loss and a tangent hyperbolic (TanH) activation function \cite{dubey_activation_2022} as last layer in generator, and normalisation between -1 and 1 for both the ground truth masks and the original images; (ii) with a Dice loss and a Sigmoid \cite{dubey_activation_2022} applied to the last layer. For the UNet,UNet++,DeepLabV3+ and FPN, the input data was normalised using mean and standard deviation specific to the pre-trained encoder employed. Also, these models were trained using Dice loss function.

\section{Experimental setup}
The UNet, UNet++, DeepLabV3+ and FPN segmentation networks were implemented using pytorch and an established segmentation library \cite{Yakubovskiy:2019} . The library offers a wide range of architectures and pre-trained encoders. We initialised the network with weights pre-trained on Imagenet \cite{deng_imagenet_2009}. We trained the networks using Adam optimizer \cite{kingma_adam_2017} with a learning rate of 0.001 and a batch size of 16. 
We implemented early stopping based on Dice loss \cite{jadon_survey_2020} on the validation set, with a maximum number of epochs of 1000 and a patience of 20. The pix2pix model was implemented similarly to \cite{pix2pix2017} . Prior to feeding inputs to the model, images were concatenated with their ground truth masks. The overall loss used to train the pix2pix model comprised a weighted summation of the cGAN loss\cite{mirza_conditional_2014} and a mean absolute error loss (L1) \cite{janocha_loss_2017}, where the latter was multiplied by 100. Both the generator and the discriminator were trained with an Adam optimizer with a learning rate of 0.0002 and a beta value of 0.999. Input data was augmented online during training using dropout and random flipping. All models were trained and validated across three folds of cross-validation. The best-performing models from the cross-validation experiments, for each segmentation network investigated, were then evaluated on the test set. To verify generalisability, we tested all networks on an additional dataset containing images of patient with titin mutation \cite{parker_affimers_2023}. 
All networks were trained on a workstation with a 12th Gen Intel(R) Core(TM) i7-12700H CPU and NVIDIA T600 GPU. Metrics used to evaluate the performance of all segmentation models included the Dice score and pixel-wise recall (sensitivity) \cite{minaee_image_2020}.

\begin{figure} [ht]
   \begin{center}
   \begin{tabular}{c} 
   \includegraphics[height=6cm]{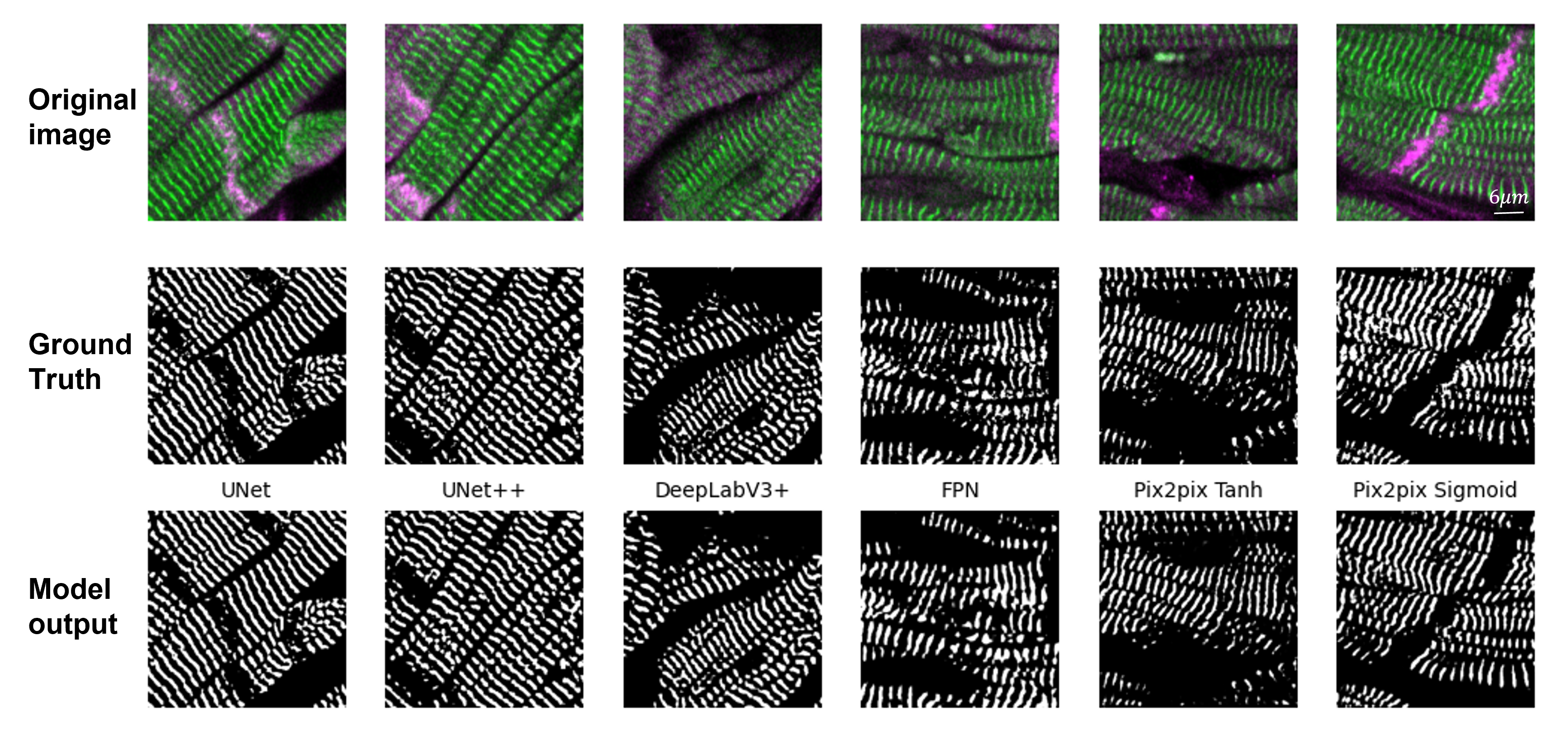}
   \end{tabular}
   \end{center}
   \caption[example] 
   { \label{fig:zdisks} 
Original images, ground truths and segmentations of cardiomyocyte Z-disks imaged with confocal microscopy obtained with semi-interactive labelling in ilastik.}
   \end{figure}
\begin{table}[ht]
\scriptsize{
\caption{ Performance of segmentation models on validation (3-fold), test and external data sets.} 
\label{tab:Results}
\begin{center}       
\begin{tabular}{|l|l|l|l|l|l|l|l|} 
\hline
\rule[-1ex]{0pt}{3.5ex}  \textbf{Model} &  \textbf{Dice (CVal)} & \textbf{Recall (CVal)} &  \textbf{Dice (Test)}  & \textbf{Recall (Test)} & \textbf{Dice (Ext)} & \textbf{Recall (Ext)}  \\
\hline
\rule[-1ex]{0pt}{3.5ex}  UNet & 0.90/0.90/0.90  & 0.90/0.89/0.91 & 0.90 & 0.91 & 0.76 & 0.74 \\
\hline
\rule[-1ex]{0pt}{3.5ex}  UNet++ & \textbf{0.92/0.92/0.92}  & 0.92/0.94/\textbf{0.94} & \textbf{0.91} & \textbf{0.93} & 0.74 & 0.75   \\
\hline
\rule[-1ex]{0pt}{3.5ex}  DeepLabV3+  & 0.84/0.85/0.88   & 0.85/0.87/0.91  & 0.85 & 0.86 & 0.62 & 0.64 \\
\hline
\rule[-1ex]{0pt}{3.5ex}  FPN & 0.82/0.83/0.83  & 0.83/0.87/0.84  &  0.83 & 0.86  & 0.55 & 0.55\\  
\hline
\rule[-1ex]{0pt}{3.5ex}  Pix2pix (TanH) & 0.81/0.80/0.74 & 0.81/0.78/0.80 & 0.83 & 0.80 & \textbf{0.77} & 0.79 \\  
\hline
\rule[-1ex]{0pt}{3.5ex}  Pix2pix (sigmoid) & 0.70/0.66/0.72 & \textbf{0.93}/\textbf{0.97}/0.90 & 0.76 & 0.90 & 0.69 & \textbf{0.91} \\  
\hline
\end{tabular}
\end{center}
}
\end{table}

\section{Results and Discussion}

We trained and validated our networks across 3-fold cross-validation basis and evaluated the best performing model for each fold on an independent test set, as well as an external dataset (Tab.~\ref{tab:Results}). On the test set, UNet++ performs best with an average Dice score of 0.91 and a recall score of 0.93. 
Although producing realistic results (Fig.~\ref{fig:zdisks}), the cGAN pix2pix models did not outperform the other segmentation networks investigated, across the cross-validation experiments or on the independent test set.
To assess generalisability, we evaluated all trained models on an external dataset, consisting of images from a patient with titin mutation. We notice a significant drop in performance across both metrics compared to the performance on the independent test set for the established segmentation networks. Possible explanations are: training dataset was limited to images from one patient, which is not representative enough. There are many sources of variability across different patient acquisitions: staining, microscope settings, brightness, morphology of the sample (trained on healthy, tested on mutation). Increasing the size and diversity of training data would help improve model robustness and generalisability. Conversely, we found that the pix2pix models outperformed all other segmentation networks on the external dataset, demonstrating its improved generalisation capacity relative to the latter. 
The segmentation masks obtained will be used in future work, to derive morphological measurements of Z-disks (e.g. width, length, curvature, etc.) and identify differences between normal/healthy cardiac tissue and disease-related changes.

\section{Conclusion}
Accurated segmentation of the Z-disks in cardiomyocyte images is essential for quantifying disease-related changes to cardiac microstructure. In this study, we evaluated the performance of several automated, deep learning-based segmentation approaches for delineating Z-disks in cardiomyocyte confocal microscopy images. 
We observe that all methods are able to segment Z-disks with Dice scores above 76\%. While UNet++ achieved the best segmentation results across the cross-validation experiments and on the independent test set, pix2pix was able to generalise the best to external data comprising images from a different patient with titin mutation. 
To improve generalisability, larger, more representative datasets are required, which will be the subject of future work. 
\acknowledgments 
This work was done as part of an Mres research project within the AI for Medical Diagnosis and Care CDT funded by UKRI. We confirm that the work was not submitted for publication or presentation elsewhere.

\bibliographystyle{spiebib} 
\bibliography{report} 

\end{document}